\documentclass[twocolumn]{aastex631}

\usepackage{CJKutf8}
\usepackage{physics}

\newcommand{\SigmaSFR}{\Sigma_\mathrm{SFR}}
\newcommand{\Sigmagas}{\Sigma_\mathrm{gas}}
\newcommand{\Sigmamol}{\Sigma_\mathrm{mol}}
\newcommand{\Sigmaatom}{\Sigma_\mathrm{atom}}
\newcommand{\vcirc}{v_\mathrm{circ}}
\newcommand{\SFEorb}{\epsilon_\mathrm{orb}}
\newcommand{\SFEorbsub}[1]{\epsilon_\mathrm{orb,\,#1}}
\newcommand{\alphaCC}{\alpha_\mathrm{CC}}
\newcommand{\alphaCCsub}[1]{\alpha_\mathrm{CC,\,#1}}

\begin{document}

\title{The Impact of Shear on Disk Galaxy Star Formation Rates}

\author{Xena Fortune-Bashee}
\affiliation{Department of Astronomy, University of Virginia, 530 McCormick Road, Charlottesville, VA 22904, USA}

\author[0000-0003-0378-4667]{Jiayi Sun \begin{CJK*}{UTF8}{gbsn}(孙嘉懿)\end{CJK*}}
\altaffiliation{NASA Hubble Fellow}
\affiliation{Department of Astrophysical Sciences, Princeton University, 4 Ivy Lane, Princeton, NJ 08544, USA}

\author[0000-0002-3389-9142]{Jonathan C. Tan}
\affiliation{Department of Astronomy, University of Virginia, 530 McCormick Road, Charlottesville, VA 22904, USA}
\affiliation{Dept. of Space, Earth \& Environment, Chalmers University of Technology, Gothenburg, Sweden}

\begin{abstract}
Determining the physical processes that control galactic-scale star formation rates is essential for an improved understanding of galaxy evolution. The role of orbital shear is currently unclear, with some models expecting reduced star formation rates (SFRs) and efficiencies (SFEs) with increasing shear, e.g., if shear stabilizes gas against gravitational collapse, while others predicting enhanced rates, e.g., if shear-driven collisions between giant molecular clouds (GMCs) trigger star formation. Expanding on the analysis of 16 galaxies by Suwannajak, Tan, \& Leroy (2014), we assess the shear dependence of SFE per orbital time ($\epsilon_\mathrm{orb}$) in 49 galaxies selected from the PHANGS-ALMA survey. In particular, we test a prediction of the shear-driven GMC collision model that $\epsilon_\mathrm{orb}\propto(1-0.7\beta)$, where $\beta\equiv{d}\:\mathrm{ln}\:v_\mathrm{circ}/d\:\mathrm{ln}\:r$, i.e., SFE per orbital time declines with decreasing shear. We fit the function $\epsilon_\mathrm{orb}=\epsilon_\mathrm{orb,\,0}(1-\alpha_\mathrm{CC}\beta)$ finding $\alpha_\mathrm{CC}\simeq0.76\pm0.16$; an alternative fit with $\epsilon_\mathrm{orb}$ normalized by the median value in each galaxy yields $\alpha_\mathrm{CC}^*=0.80\pm0.15$. These results are in good agreement with the prediction of the shear-driven GMC collision theory. We also examine the impact of a galactic bar on $\epsilon_\mathrm{orb}$ finding a modest decrease in SFE in the presence of bar, which can be attributed to lower rates of shear in these regions. We discuss the implications of our results for the GMC life cycle and environmental dependence of star formation activity.
\end{abstract}

\section{Introduction}\label{sec:intro}

The SFR of galactic disks is of fundamental importance to the evolution of these systems and to galaxy evolution in general. Correlations between SFR, gas content, and galactic dynamical properties have been found on both global and local, i.e., $\lesssim\:$kpc, scales \citep[e.g.,][]{2012ARA&A..50..531K,2024arXiv240319843S}. These are typically expressed in the form of ``star formation laws''. For example, the Kennicutt-Schmidt (KS) law takes the form $\SigmaSFR\propto\Sigmagas^{\alpha_g}$, where $\SigmaSFR$ is the disk plane surface density of SFR and $\Sigmagas$ is the total (atomic + molecular) gas mass surface density. Superlinear values of $\alpha_g\simeq1.5$ have been found \citep[e.g.,][]{1998ApJ...498..541K,2012ARA&A..50..531K}.

The ``Dynamical KS'' law \citep[or ``Silk-Elmegreen'' law;][]{1997RMxAC...6..165E,1997ApJ...481..703S} takes the form 
\begin{equation}
\SigmaSFR=B_\Omega\Sigmagas\Omega,\label{eq:dynamicalks}
\end{equation}
where $\Omega$ is the galactic orbital angular frequency and $B_\Omega$ is a dimensionless normalization constant. \citet{1998ApJ...498..541K} found this was an equally good description of a sample of disk-averaged measurements of galactic disks and circumnuclear starbursts \citep[see also, e.g.,][]{Leroy_etal_2008,Tan_2010,Sun_etal_2023}. Such a dynamical KS law has been proposed to arise naturally if large scale galactic dynamical processes, such as spiral arm passage \citep[e.g.,][]{1989ApJ...339..700W} or growth of large-scale instabilities \citep[e.g.,][]{1991ApJ...378..139E}, are the rate limiting step for star formation.

However, considering that most star formation occurs in highly clustered $\sim 1{-}10$~pc regions within GMCs, \citet{Tan_2000} proposed shear-driven GMC-GMC collisions in a Toomre $Q\sim1$, flat rotation curve disk could also explain the dynamical KS law. The key prediction of this theory is that $\SigmaSFR$ is related to the galactic orbital period and shear via\footnote{Note that Eq.~\ref{eq:pred} was derived via a first order expansion in $\beta$ (i.e., valid for $\beta\lesssim0.5$).}
\begin{equation}
\SigmaSFR = B_\Omega\,\Sigmagas\,\Omega\,(1-0.7\beta),\label{eq:pred}
\end{equation}
where $\beta\equiv\dd\ln{\vcirc}/\dd\ln{r}$ is the logarithmic derivative of the rotation curve $\vcirc(r)$, i.e., taking the value $\beta=1$ in the flat rotation curve case. Higher $\beta$ values, closer to the solid body rotation case of $\beta=1$, imply lower orbital shear strength, consistent with alternative formulations with Oort's constants \citep[see, e.g.,][]{Binney_Tremaine_1987}.

\citet{Tan_2010} carried out a test of the shear-driven GMC law, along with five other star formation laws, against a sample of 12 nearby, resolved disk galaxies, focusing on molecular-rich regions, i.e., with $\Sigma_\mathrm{H2}\gtrsim\Sigma_\mathrm{HI}$. The shear-driven law (Eq.~\ref{eq:pred}) was found to be more favored than the simple dynamical KS law (Eq.~\ref{eq:dynamicalks}), although two other SF laws, i.e., a ``molecular KS'' law with $\Sigma_{\rm SFR}$ proportional to molecular gas mass surface density and the ``turbulence-regulated'' law of \citet{2009ApJ...699..850K}, gave similarly good fits to the data. \citet{Suwannajak_etal_2014} followed up this study by examining 16 galaxies, demonstrating with high significance that higher shear rates correlate with higher SFEs per local orbital time, $\SFEorb$. A modestly higher (by a factor of 1.3) average value of $\SFEorb$ was found in the sub-sample of barred galaxies compared to the non-barred sample. However, this modest enhancement was not statistically significant given the small sample sizes.

Here we test the prediction of Eq.~(\ref{eq:pred}) with higher quality data for a sample of 49 galaxies. We present our methods in \S\ref{sec:method}, our results in \S\ref{sec:results}, and a summary and discussion in \S\ref{sec:discuss}.

\section{Method} \label{sec:method}

We test Eq.~\ref{eq:pred} using observations of $\SigmaSFR$, $\Sigmagas$, $\Omega$, and $\beta$ in 49 galaxies from the PHANGS--ALMA survey \citep{Leroy_etal_2021a}.
We extract these measurements from machine-readable tables \citep[version~4.0;][]{Sun_etal_2022}, reported as azimuthal averages in 500~pc-wide radial bins across each galaxy (see \S3.1 therein). Here we briefly describe how these quantities were derived and refer readers to \citet{Sun_etal_2022} for more details.
Values of $\SigmaSFR$ were derived by combining ground-based H$\alpha$ narrow-band and WISE 22~$\mu$m imaging data, following the calibration by \citet{Belfiore_etal_2023}.
$\Sigmamol$ was derived from PHANGS--ALMA CO\,(2--1) observations \citep{Leroy_etal_2021a} using a metallicity-dependent CO-to-H$_2$ conversion factor \citep[][]{Sun_etal_2020a,Sun_etal_2020b}.
$\Sigmaatom$ was derived from a compilation of literature and new H\textsc{i} 21~cm observations, including THINGS \citep{Walter_etal_2008}, VIVA \citep{Chung_etal_2009}, VLA--HERACLES \citep{leroy_etal_2012}, and PHANGS--VLA (PI: D.~Utomo). All these surface density measurements are corrected for galaxy inclination.
Lastly, $\vcirc$ and $\beta$ were based on CO kinematics \citep{Lang_etal_2020}, or more specifically, smooth functional fits to the observed CO rotation curves \citep[E.~Rosolowsky, priv.~comm.; see detailed descriptions in][]{Sun_etal_2022}.

We use these measurements to test Eq.~\ref{eq:pred} by recasting it into the following form \citep[also see][]{Tan_2010,Suwannajak_etal_2014}:
\begin{equation}
\SFEorb\equiv\SigmaSFR\,t_\mathrm{orb}\,/\,\Sigmagas=2\pi{B}_\Omega\,(1-0.7\beta),\label{eq:SFEorb_pred}
\end{equation}
where $t_\mathrm{orb}=2\pi/\Omega$ is orbital period and $\SFEorb$ is dimensionless SFE per orbit.
We derive $\SFEorb$ and $\beta$ for each radial bin in each galaxy and assess if their relationship can be well described by Equation~\ref{eq:SFEorb_pred}.

We note that the original formulation of the \citet{Tan_2000} model considers total gas mass surface density, $\Sigmagas$, as the relevant parameterization of the gas reservoir.
Nonetheless, recent works find $\SigmaSFR$ correlates more strongly with $\Sigmamol$, suggesting a tighter link between star formation and molecular gas \citep[e.g.,][]{Bigiel_etal_2008,Leroy_etal_2008,Schruba_etal_2011}.
Besides, the limited availability of H\textsc{i} 21~cm data for the PHANGS--ALMA galaxy sample \citep[see][for a compiled list]{Sun_etal_2022} means we can only derive $\Sigmagas=\Sigmamol+\Sigmaatom$ for a subsample of 32 galaxies, whereas we have $\Sigmamol$ measurements for 49.
Given these considerations, we use two operational definitions for SFE per orbit:
\begin{align}
\SFEorb &\equiv \SigmaSFR\,t_\mathrm{orb}\,/\,\Sigmagas, \label{eq:SFEorb_tot}\\
\SFEorbsub{mol} &\equiv \SigmaSFR\,t_\mathrm{orb}\,/\,\Sigmamol.\label{eq:SFEorb_mol}
\end{align}
\noindent We derive each of these for as many galaxies and radial bins as allowed by data availability and examine their relationships with $\beta$. 

\begin{figure}[ht!]
\epsscale{1.1}
\hspace{-0.02\textwidth}\plotone{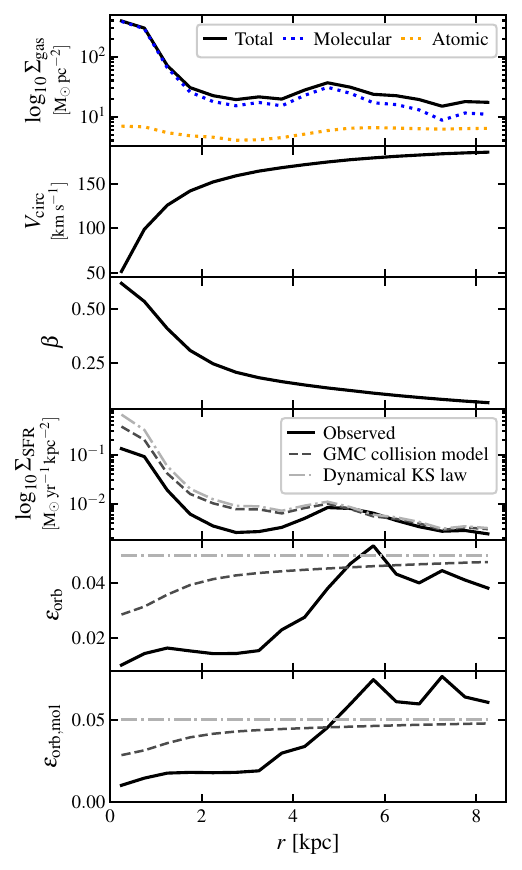}
\vspace{-0.8\baselineskip}
\caption{Example dataset for NGC~4321. Top panel shows radial profile of mass surface density of total (black solid), molecular (blue dotted), and atomic gas (orange dotted). Second and third panels show orbital velocity and its logarithmic derivative, $\beta$. Fourth panel shows the observed SFR surface density profile (solid) and theoretical predictions of the \citet{Tan_2000} shear-driven GMC collision model, $\Sigma_\mathrm{SFR}=B_\Omega\Sigma_\mathrm{gas}\Omega(1-0.7\beta)$ (dashed), and the dynamical KS model \citep{1997RMxAC...6..165E,1997ApJ...481..703S} with a fixed SFE per orbit, i.e., $\Sigma_\mathrm{SFR}=B_\Omega\Sigma_\mathrm{gas}\Omega$ with $B_\Omega=8\times10^{-3}$ (dotted). Fifth panel shows observed SFE per orbit with reference to the total gas as well as theoretically predicted values (same models as in the fourth panel). Sixth panel is the same as fifth, but now showing SFE per orbit with reference to the molecular gas.
\label{fig:radial_profile}}
\end{figure}

\section{Results}\label{sec:results}

\subsection{Impact of Shear on Star Formation}\label{sec:betatest}

Figure~\ref{fig:radial_profile} shows example data for galaxy NGC~4321. In particular, the fourth panel shows the observed SFR surface density compared to the SFR surface density predicted by the shear-driven GMC collision model (Eq.~\ref{eq:pred}) with $B_\Omega=8.0\times10^{-3}$, which corresponds to a flat rotation curve ($\beta=0$) limiting value of $\SFEorb=2\pi{B}_\Omega=0.05$. The prediction of the simpler dynamical KS model (Eq.~\ref{eq:dynamicalks}) with $\SFEorb$ independent of $\beta$, is shown by the dotted line. The fifth panel shows the observed SFE per orbit with reference to total gas, $\SFEorb$, and the predictions from the above two theoretical models. The sixth panel shows the equivalent results for SFE per orbit with reference to molecular gas, $\SFEorbsub{mol}$. The data shown in Fig.~\ref{fig:radial_profile} indicate that there is a reduced SFE per orbit in regions of lower shear, i.e., higher $\beta$.

\begin{figure*}[htb]
\centering
\epsscale{1.2}
\plotone{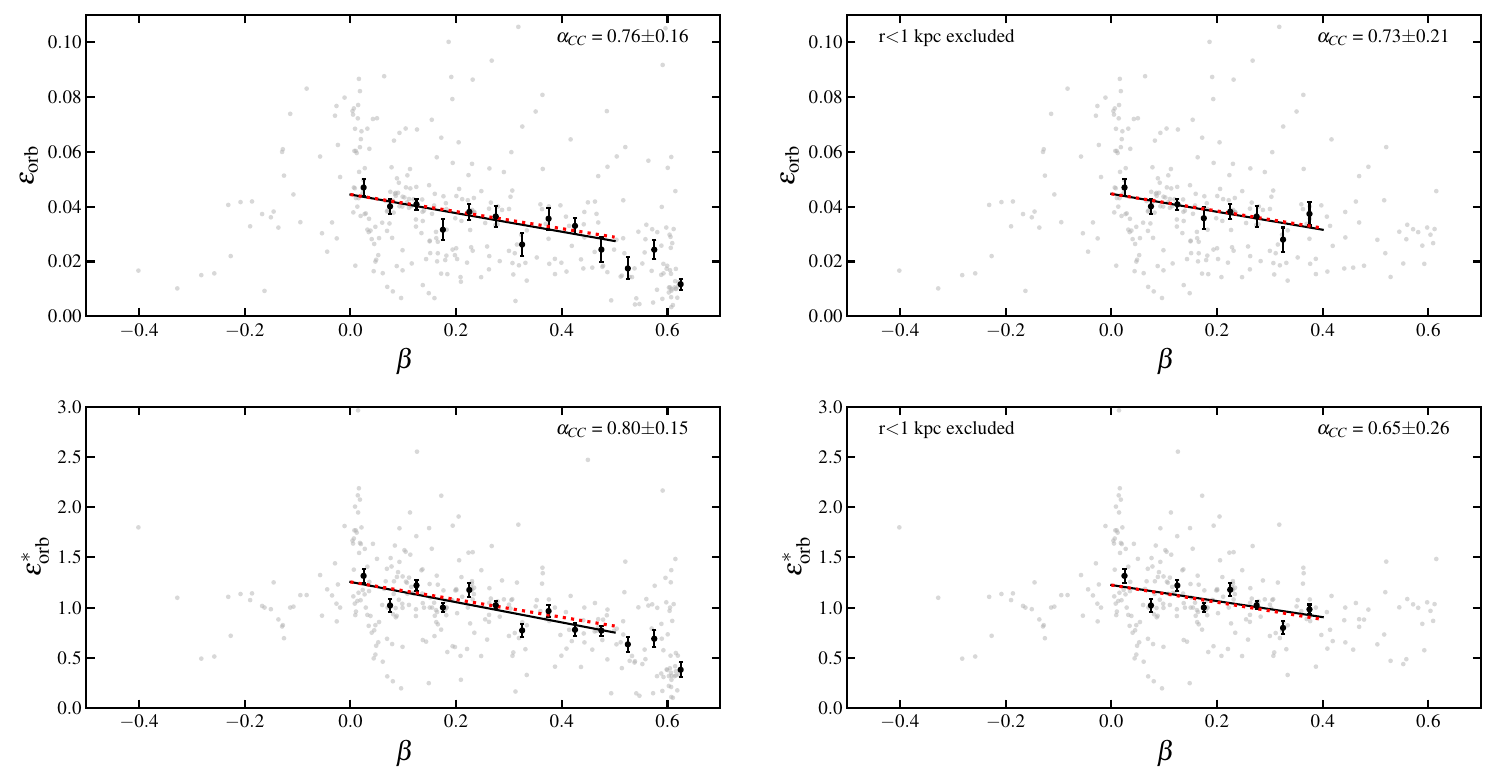}
\vspace{-1.5\baselineskip}
\caption{{\it (a) Top left:} Total SFE per orbit, $\SFEorb$, versus rotation curve gradient, $\beta$, for 321~annuli located in 32 PHANGS-ALMA galaxies with observations probing molecular and atomic gas (grey points). Black points show the median values of $\SFEorb$ in bins of $\Delta\beta=0.05$ that have at least 10 data points, with associated error bars showing the uncertainties in the median. The solid black line shows the best-fit function of the form $\SFEorb=\SFEorbsub{0}(1-\alphaCC\beta)$ to the binned medians with $\beta\leq0.5$, with $\alphaCC$ reported in the upper right corner of each panel. The red dotted line shows the equivalent fit with a fixed $\alphaCC=0.7$, i.e., the prediction of the shear-driven GMC collision model.
{\it (b) Top right:} As (a), but now excluding annuli within $r=1\:$kpc.
{\it (c) Bottom left:} As (a), but with $\SFEorb$ normalised by the median value of each galaxy (i.e., ``galaxy-normalised'' $\SFEorb^*$).
{\it (d) Bottom right:} As (c), but now excluding annuli within $r=1\:$kpc.
A trend of declining efficiency with increasing $\beta$ is seen in all cases.
\label{fig:general2}}
\end{figure*}

In Figure \ref{fig:general2}a we plot
$\SFEorb$ versus $\beta$ across 321 independent annuli in 32 PHANGS-ALMA galaxies. We evaluate median values of $\SFEorb$ in bins of uniform width $\Delta\beta=0.05$, excluding bins with $<10$ data points. These binned medians are shown with black points, along with their uncertainties.

Next, for all binned medians with $\beta\leq0.5$\footnote{This choice, following \citet{Suwannajak_etal_2014}, avoids the regime where $\beta$ is no longer significantly smaller than unity, where Eq.~\ref{eq:pred} is no longer a good approximation (see \S\ref{sec:intro}).}, we fit a function:
\begin{equation}
\SFEorb=\SFEorbsub{0}(1-\alphaCC\beta),\label{eq:predepsorb}
\end{equation}
where $\SFEorbsub{0}$ is the value of $\SFEorb$ in the flat rotation curve limit of $\beta=0$. The results are $\SFEorbsub{0}=0.045\pm0.002$ and $\alphaCC=0.76\pm0.16$, shown by the solid black line in Fig.~\ref{fig:general2}a. These fit parameters are insensitive to choices of bin width in $\beta$ and consistent with alternative, non-parametric methods for reconstructing the median trends. An equivalent fit in which $\alphaCC$ is fixed at 0.7, i.e., the predicted value of the shear-driven GMC collision model, is shown by the red dotted line. We see that the observed dependence of $\SFEorb$ versus $\beta$ agrees well with the theoretical prediction.

Next, we repeat this analysis, but excluding the innermost 1~kpc of each galaxy (Fig.~\ref{fig:general2}b). These regions have the most uncertain rotation curve measurements (e.g., due to significant non-circular motions) and thus values of $\beta$. They may also have the largest systematic uncertainties in gas masses and SFRs. These excluded regions tend to have high values of $\beta$ and so the remaining data are now insufficient to measure median values for $\beta>0.4$. Nevertheless, the impact on the derived parameters appears minimal, with $\SFEorbsub{0}=0.045\pm0.002$ and $\alphaCC=0.73\pm0.21$.

Next, in Fig.~\ref{fig:general2}c, we consider ``galaxy-normalized efficiencies'', i.e., $\epsilon^*_\mathrm{orb}\equiv\SFEorb/\overline\SFEorb$, where $\overline\SFEorb$ is the galaxy-wide median value of $\SFEorb$, i.e., among all annuli of a given galaxy. This method, following \citet{Suwannajak_etal_2014}, gives an extra degree of freedom in the normalization of each galaxy and so yields a more accurate measure of the relative dependence of $\SFEorb$ with $\beta$ for the global sample. We fit the function:
\begin{equation}
\SFEorb^*=\SFEorbsub{0}^*(1-\alphaCC^*\beta),
\end{equation}
and derive $\alphaCC^*=0.80\pm0.15$. Figure~\ref{fig:general2}d presents the equivalent results when the innermost 1~kpc regions are excluded, for which $\alphaCC^*=0.65\pm0.26$.

We conclude there is a significant trend detected of declining $\SFEorb$ with increasing $\beta$, i.e., reduced SFE per orbit as rate of shear decreases. Furthermore, the functional form of this trend agrees well with the prediction of the shear-driven GMC collision theory of \citet{Tan_2000}. The result is insensitive to whether or not the innermost annuli are excluded or whether galaxy-normalized quantities are used. Thus our results are evidence in favor of shear-driven GMC collisions being the primary mechanism regulating star formation in galactic disks.

Our results agree with those of \citet{Suwannajak_etal_2014}, who found $\alphaCC=1.13\pm0.49$ for their full dataset based on older observations for 16 galaxies (which overlaps partially, 7/16, with our sample), or $\alphaCC=1.10\pm0.44$ when innermost annuli were excluded. For galaxy-normalized data, \citet{Suwannajak_etal_2014} found $\alphaCC^*=1.39\pm0.32$ and $\alphaCC^*=1.35\pm0.31$ for the full and inner-kpc excluded datasets, respectively. Their larger values of $\alphaCC$ and $\alphaCC^*$ could be related to systematic differences in analysis methods. In particular, we use a more realistic, metallicity-dependent conversion factor of CO line luminosity to molecular gas mass, which tends to reduce our gas masses in inner, high-$\beta$ regions, thus causing $\SFEorb$ to increase in this regime and thus reducing our value of $\alphaCC$ compared to that of \citet{Suwannajak_etal_2014}.

\begin{figure*}[ht!]
\epsscale{1.2}
\plotone{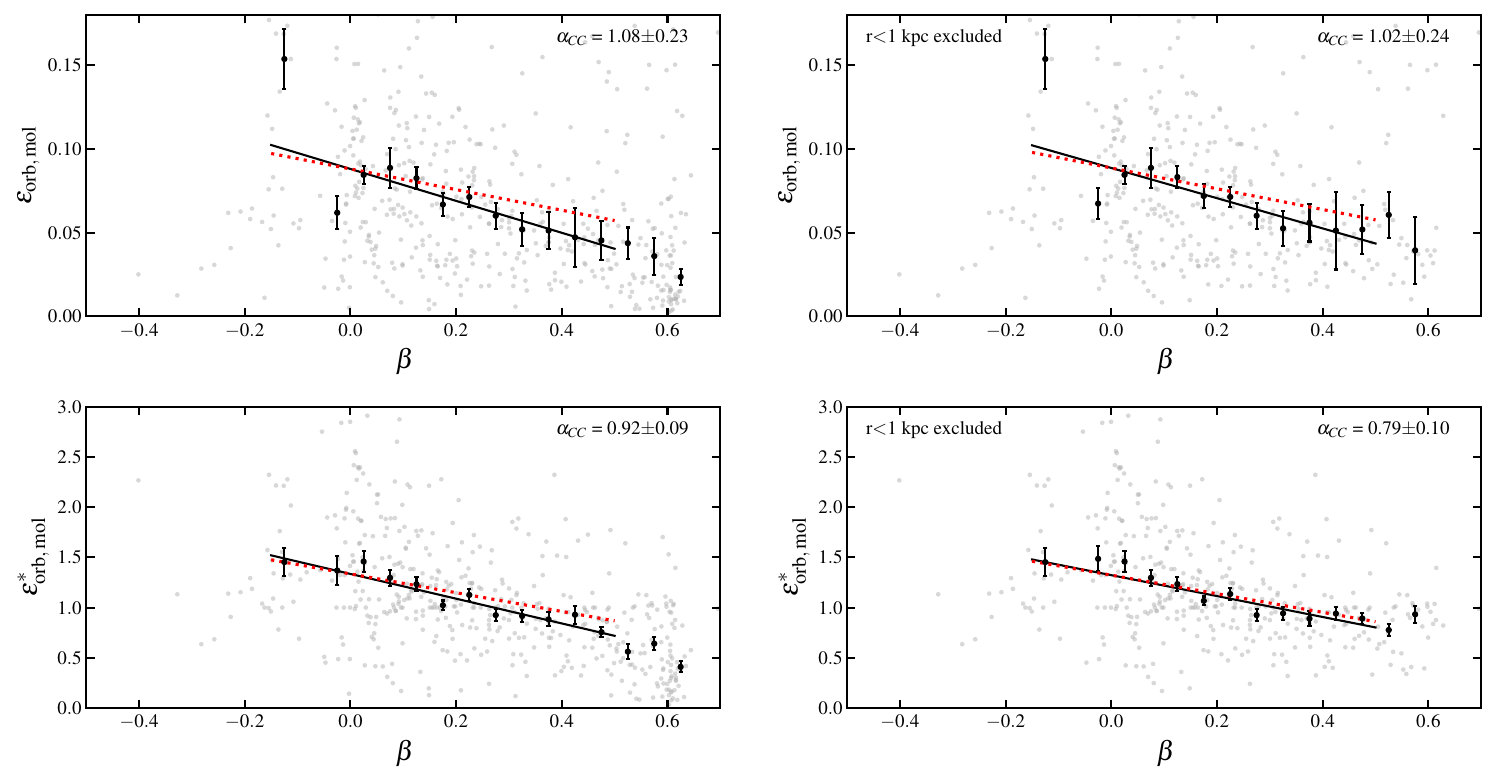}
\vspace{-1.5\baselineskip}
\caption{{\it (a) Top left:} SFE from molecular gas per orbit, $\SFEorbsub{mol}$, versus rotation curve gradient, $\beta$, for 489~annuli located in 49 PHANGS-ALMA galaxies (grey points). Median values of $\SFEorbsub{mol}$ in bins of $\Delta\beta=0.05$ that have $\geq10$ data points are shown by black points, with uncertainties shown by the error bars. The best-fitting function of the form $\SFEorbsub{mol}=\SFEorbsub{mol,\,0}(1-\alphaCCsub{mol}\beta)$ to median values with $\beta\leq0.5$ is shown by solid black line, with $\alphaCCsub{mol}$ reported in upper right. Red dotted line shows the equivalent fit when $\alphaCCsub{mol}=0.7$.
{\it (b) Top right:} As (a), but now excluding annuli within $r=1\:$kpc.
{\it (c) Bottom left:} As (a), but now showing the ``galaxy-normalized efficiency'', $\SFEorbsub{mol}^*$.
{\it (d) Bottom right:} As (c), but now excluding annuli within $r=1\:$kpc.
\label{fig:general3}}
\end{figure*}

In Figure \ref{fig:general3} we repeat the above analysis, but now considering $\SFEorbsub{mol}$ across 489 independent annuli in 49 PHANGS-ALMA galaxies. We find $\SFEorbsub{mol,\,0}=0.088\pm0.005$ and $\alphaCCsub{mol}=1.08\pm0.23$. Excluding the innermost 1~kpc has only a minor effect, with $\alphaCCsub{mol}=1.02\pm0.24$. For the galaxy-normalized data, we find $\alphaCCsub{mol}^*=0.92\pm0.09$. Excluding the innermost 1~kpc yields $\alphaCCsub{mol}^*=0.79\pm0.10$. These slightly larger values of $\alphaCC$ in the molecular case are likely caused by the atomic gas making up a larger fraction of the total in the outer regions that tend to have smaller values of $\beta$. Thus SFE per orbit with respect to molecular gas is larger in these low-$\beta$ regions, resulting in a steeper gradient in the $\SFEorbsub{mol}$ versus $\beta$ plane.

In summary, the results of Fig.~\ref{fig:general3} indicate a clear decrease in SFE per orbit with increasing $\beta$ when only considering the molecular gas, similar to our conclusion for the total gas.
 
\subsection{Impact of a Bar on Star Formation}

\begin{figure*}[ht!]
\epsscale{1.2}
\plotone{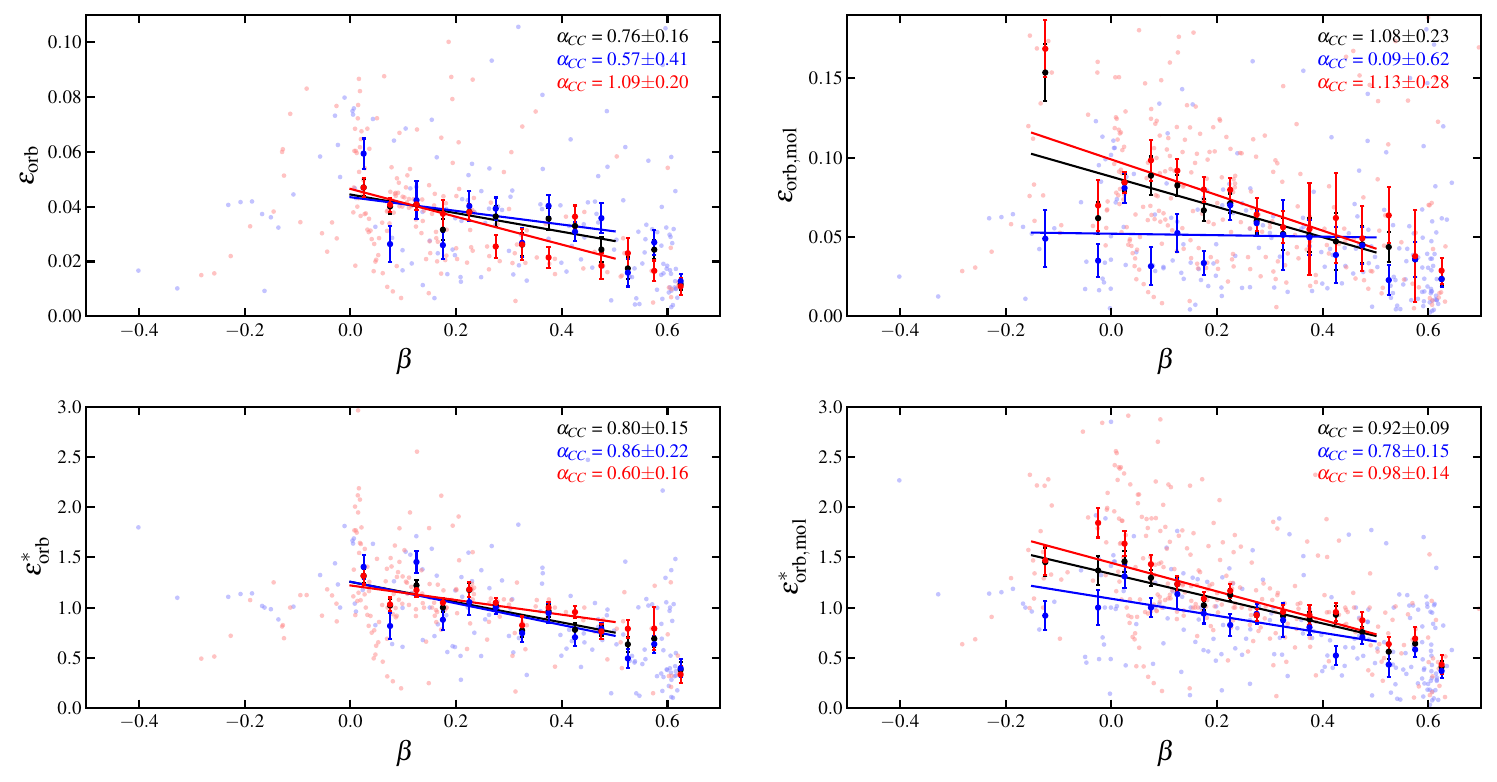}
\vspace{-1.5\baselineskip}
\caption{{\it (a) Top left:} Similar to Fig.~\ref{fig:general2}a, total SFE per orbit, $\SFEorb$, versus rotation curve gradient, $\beta$, for 32 PHANGS-ALMA galaxies with molecular and atomic gas data. Barred/non-barred annuli are shown in blue/red, respectively. Median values of $\SFEorb$ of the combined sample in bins of $\Delta\beta=0.05$ with $\geq10$ data points are shown by black points. In this range of $\beta$ we also show medians of barred/non-barred annuli with dark blue/red points, respectively. Best-fitting functions to the medians of form $\SFEorb=\SFEorbsub{0}(1-\alphaCC\beta)$ are shown by the black, blue and red lines for total, barred and non-barred samples, respectively.
{\it (b) Top right:} As (b), but now for SFE from molecular gas per orbit, $\SFEorbsub{mol}$, versus $\beta$, for 49 PHANGS-ALMA galaxies.
{\it (c) Bottom left:} As (a), but now showing ``galaxy-normalized efficiency'', $\SFEorb^*$. 
{\it (d) Bottom right:} As (b), but now showing $\SFEorbsub{mol}^*$. 
\label{fig:general4}
}
\end{figure*}

To test the effect of a stellar bar on $\SFEorb$, we divide the measurements into two subsamples, i.e., annuli that do and do not overlap with the footprint of a stellar bar, \citep[as characterized by][also see \citealt{Sun_etal_2022}]{Querejeta_etal_2021}. We note that this comparison only measures the potential impact of a bar on SFE assessed in global annular averages. Overall, for regions with both atomic and molecular data, the barred/non-barred samples consist of 143/178 annuli, respectively. For the SFE from molecular gas analysis, barred/non-barred samples consist of 194/295 annuli, respectively.

Figure~\ref{fig:general4} shows these data, distinguishing between barred/non-barred data. We evaluate median values of $\SFEorb$ of these subsamples. The barred sample has $\overline{\SFEorb}=0.0324\pm0.0017$ and the non-barred has $\overline{\SFEorb}=0.0375\pm0.0015$. Thus, the barred sample shows a reduced SFE relative to the non-barred by a factor of 0.86. The median values of $\beta$ of the barred and non-barred subsamples are $0.364\pm0.022$ and $0.117\pm0.012$, respectively. For these values of $\beta$ and for $\alphaCC=0.76$, Eq.~(\ref{eq:predepsorb}) predicts that the barred sample would have lower SFE per orbit than the non-barred sample by a factor of 0.79. Thus the differences we see between the barred and non-barred samples are attributable to the difference in their $\beta$ ranges, i.e., lower shear of barred regions explains their reduced SFE per orbit.

We repeat the above analysis, but now considering $\SFEorbsub{mol}$. Evaluating median values for $\SFEorbsub{mol}$ of these subsamples we find $\overline{\SFEorbsub{mol}}=0.0425\pm0.0025$ for the barred sample and $\overline{\SFEorbsub{mol}}=0.0780\pm0.0039$ for the non-barred. Again, presence of a bar is associated with reduced SFE per orbit, now by a factor of 0.54. The median values of $\beta$ of these barred/non-barred sub-samples are $0.370\pm0.021$ / $0.165\pm0.013$. For these values of $\beta$ and for $\alphaCCsub{mol}=1.08$, Eq.~(\ref{eq:predepsorb}) predicts that the barred sample would have lower SFE per orbit than the non-barred sample by a factor of 0.73. Thus most of the observed reduction is attributable to the lower shear of the barred regions, but with a tentative hint of an additional suppression of SFE in those regions.

To further investigate the dependence of SFE with shear in the barred/non-barred samples and compare with fitted parameters of \S\ref{sec:betatest}, we again turn to binned statistics and focus on data with $\beta<0.5$. In Fig.~\ref{fig:general4} we show fits for the barred/non-barred subsamples across the same set of $\beta$ bins used for fitting the global trend. This conserves the information content of the original sample. The barred sample has $\SFEorbsub{0}=0.043\pm0.007$ and $\alphaCC=0.57\pm0.41$, while the non-barred sample has $\SFEorbsub{0}=0.047\pm0.003$ and $\alphaCC=1.09\pm0.20$. For the galaxy-normalized data, the barred sample has $\alphaCC^*=0.86\pm0.22$, while the non-barred sample has $\alphaCC^*=0.60\pm0.16$. We see that barred and non-barred samples have best-fit parameters that are consistent with each other within the uncertainties.

When considering $\SFEorbsub{mol}$, the barred sample has $\SFEorbsub{mol,\,0}=0.052\pm0.008$ and $\alphaCCsub{mol}=0.09\pm0.62$, while the non-barred sample has $\SFEorbsub{mol,\,0}=0.099\pm0.007$ and $\alphaCCsub{mol}=1.13\pm0.28$. For the galaxy-normalized data, the barred sample has $\alphaCCsub{mol}^*=0.78\pm0.15$, while the non-barred sample has $\alphaCCsub{mol}^*=0.98\pm0.14$. From this we see the dependence of $\SFEorbsub{mol}$ on shear is greater in the non-barred sample: in particular, in the high shear regime with $\beta\simeq0$ the SFE per orbit is larger in non-barred environments. However, we caution that the significance of this effect is modest. There are also known systematic biases \citep[e.g., bar dynamics affecting rotation curve measurements, biases in SFR and gas mass estimates near galaxy centers; see][]{Lang_etal_2020,Sun_etal_2023} that impact the barred and non-barred samples differently. These systematic effects could potentially explain at least part of the apparent differences in the observed trends.

\section{Summary and Discussion} \label{sec:discuss}

We have investigated SFE per orbit, $\SFEorb$, with reference to total gas content, and how it depends on local rate of shear, as parameterized by $\beta\equiv\dd\ln{\vcirc}/\dd\ln{r}$ in 32 galaxies from the PHANGS-ALMA survey. We have found clear evidence for a decline in $\SFEorb$ as rate of shear decreases, i.e., as $\beta$ increases from near zero of the flat rotation curve regime towards $\beta\sim1$ of the solid body rotation regime. Such a trend is the opposite of that expected if growth of gravitational instabilities is the rate-limiting step for galactic disk star formation \citep[e.g.,][]{1991ApJ...378..139E}. 

Motivated by a prediction of the shear-driven GMC collision theory of \cite{Tan_2000}, we fit a function $\SFEorb=\SFEorbsub{0}(1-\alphaCC\beta)$ to the data where $\beta\leq0.5$. The shear-driven GMC collision model predicts $\alphaCC=0.7$; we find $\alphaCC=0.76\pm0.16$, or $\alphaCC=0.73\pm0.21$ when inner kpc regions are excluded. For a fit to efficiencies normalized by the median value in each galaxy, we find $\alphaCC^*=0.80\pm0.15$ and $\alphaCC^*=0.65\pm0.26$ when inner kpc regions are excluded. For SFE per orbit with respect to molecular gas, similar but slightly larger values of $\alphaCC$ are derived. Thus our primary conclusion is that the observational data are consistent with the theoretical prediction of the shear-driven GMC collision model, implying that this process may play an important, perhaps dominant, role in controlling galactic disk SFRs. 

As reviewed by \citet{2013IAUS..292...19T}, there are important implications for the evolution of GMCs and the ISM of galaxies if shear-driven GMC collisions occur at a high enough rate to trigger most star formation in galactic disks. In the shear-driven GMC collision model the formation of molecular gas, i.e., GMCs, from atomic gas is not the rate limiting step for star formation. Indeed, a significant fraction ($\gtrsim1/3$) of the disk ISM, which is self-regulated by star formation activity to have Toomre $Q\sim 1$, is assumed to be in GMCs that are relatively stable with respect to collapse due to the presence of magnetic fields and turbulence. Frequent collisions between GMCs create local regions that lose magnetic support, i.e., becoming magnetically supercritical, and collapse with relatively high efficiency to form star clusters and associations. This process then sets the clustering properties of star formation, especially making the distribution of young stars more clustered than that of molecular gas. The stochastic nature of GMC collisions leads to a broad dispersion in GMC SFEs based on instantaneous snapshots of current young star populations and associated local molecular gas, which helps explain observed distributions of efficiencies in GMCs \citep[e.g.,][]{Tan_2000,2016ApJ...833..229L}. The kinematics of these young stars are expected to be quite disturbed, which aligns with observational results from {\it Gaia} proper motion studies of Galactic young stars \citep[][]{2023ASPC..534..129W}. 

In addition, frequent shear-driven collisions between GMCs are efficient at extracting galactic orbital kinetic energy and converting it into turbulent kinetic energy in the clouds, giving additional support to much of their structures. Thus GMCs, even though containing significant amounts of locally self-gravitating gas, are typically not globally relaxed and virialized, i.e., on the largest scales a single ``GMC'' is often composed of two or more self-gravitating clouds interacting from a recent collision that overall can appear to be unbound. Another consequence of frequent collisions of GMCs is a broad distribution of angular momentum directions, including significant fractions with retrograde rotation with respect to their galaxy \citep{2009ApJ...700..358T,2018PASJ...70S..56L}, which appears relevant to observations of Milky Way \citep{2011ApJ...732...78I} and nearby galaxy \citep{2003ApJ...599..258R,2011ApJ...732...79I} GMC populations. Finally, since collisions are driven by galactic shear, this causes large-scale dense filamentary substructures of the GMCs to have preferential alignment with the galactic plane \citep{2015ApJ...805....1B}, which is observed in samples of large ($\gtrsim10\:$pc) Galactic filaments \citep[e.g.,][]{2018ApJ...864..153Z,2023A&A...675A.119G}.

Previous studies have disfavored GMC collisions as an important mechanism because of supposedly long timescales for GMCs to collide compared to estimated GMC lifetimes \citep[e.g.,][]{2007ARA&A..45..565M,Sun_etal_2022}. However, since the scale-height of molecular gas is similar to GMC sizes, their collision times need to be estimated in a 2D-geometry in a shearing disk \citep{1991ApJ...378..565G,Tan_2000,2009ApJ...700..358T,2018PASJ...70S..56L} rather than a 3D geometry \citep[e.g.,][]{2007ARA&A..45..565M}. Collision times in regions where a significant fraction ($\gtrsim1/3$) of total ISM mass is in GMCs are estimated to be ${\sim}10{-}20\%$ of a local orbital time \citep{Tan_2000,2009ApJ...700..358T,2018PASJ...70S..56L}, which is ${\sim}10{-}30\:$Myr in the regions considered in our study \citep{Sun_etal_2022}. We note that the observational estimates of GMC collision times by \citet{Sun_etal_2022}, although based on a 2D disk geometry, are still likely overestimated, since the number of clouds per unit area is limited by the data resolution and effects of gravitational focusing were not included \citep[i.e., collision cross-section should be $\sim1-2$ tidal radii, which can be larger than cloud geometric sizes; see][]{1991ApJ...378..565G,Tan_2000}. Finally, the relatively short estimates of GMC lifetimes (${\sim}5{-}30\:$Myr) based on the spatial decorrelation between CO and H$\alpha$ emission \citep[a.k.a.\ the ``tuning fork'' method; e.g.,][]{2010ApJ...722.1699S,2014MNRAS.439.3239K,2020MNRAS.493.2872C,2022MNRAS.516.3006K} have recently been called into question by \citet{2023ApJ...959....1K}, especially in molecular-rich regions, as the observed CO--H$\alpha$ decorrelation can also be reproduced if GMCs and their young stars have modest relative drift speeds of $\sim10\:\mathrm{km\:s}^{-1}$ \citep[however, see counterclaim by][]{2024arXiv240414495K}. Such drift speeds can potentially be induced by the GMC collision process itself. Thus, given the uncertainties, we consider that current estimates of GMC collision times and lifetimes are comparable and a scenario in which GMC evolution is significantly affected by frequent collisions is plausible.

Future studies are needed to further investigate and test predictions of the shear-driven GMC theory. For example, a larger set of high-resolution observations that focuses on regions of low and high shear will help to test the theory more stringently. Regions of higher shear, i.e., where the rotation curve is declining with radius, should yield a greater dynamic range in predicted SFRs and SFEs. However, if the relative speed of collisions increases in such regions, then the impact of this increased collision speed on the SFE from a given collision may need to be accounted for.

On the other hand, in low-shear regions of near solid body rotation, i.e., with $\beta\sim1$, shear-driven GMC collisions will not occur frequently enough to influence star formation and other processes must instead regulate the process. Thus, in these regions, relevant to galactic centers, late-type spiral galaxies and dwarf galaxies, one expects a potentially significant systematic differences in star formation properties, including efficiency per orbit, efficiency per local free-fall time, and the initial cluster mass function (ICMF). The results presented here are the most extensive to date focusing on the impact on $\SFEorb$ in these low-shear regions, with the overall value dropping by a factor of at least $\sim0.6$ for regions with $\beta>0.5$ compared to the flat rotation curve ($\beta=0$) regime. In terms of the impact on the ICMF, \citet{2008AJ....135..823D} investigated whether it differs in large disk galaxies compared to dwarf galaxies, finding ICMFs were indistinguishable for masses $>10^{4.4}\:M_\odot$. However, to our knowledge, an explicit investigation of the ICMF as a function of $\beta$ remains to be carried out.

Finally, there are other environments in which shear-driven GMC collisions are not expected to be important and where star formation properties may also differ from typical disk systems. These include ``overlap'' regions of colliding galaxies, e.g., in the Antennae galaxies \citep[e.g.,][]{2014ApJ...795..156W,2024MNRAS.530..597B}, although GMC collisions may still be occurring driven by flows associated with the galaxy collision. Tidal tails of interacting or stripped galaxies \citep[e.g.,][]{2019MNRAS.482.4466P} are other environments where shear-driven collisions are not expected to operate and star formation would be regulated by alternative processes. Systematic studies of samples of such sources and comparison to galactic disks is a promising avenue to better understand the processes regulating star formation across diverse galactic environments.


\vspace{\baselineskip}
{

We thank the anonymous referee and the ApJ statistics editor for their constructive feedback.
XFB acknowledges support from a Virginia Initiative on Cosmic Origins (VICO) summer undergraduate fellowship and support from the Chalmers Initiative on Cosmic Origins (CICO).
JS acknowledges support by the National Aeronautics and Space Administration (NASA) through the NASA Hubble Fellowship grant HST-HF2-51544 awarded by the Space Telescope Science Institute (STScI), operated by the Association of Universities for Research in Astronomy, Inc., under contract NAS~5-26555. JCT acknowledges support from NSF grant AST-2009674 and ERC Advanced Grant MSTAR.

This paper makes use of the following ALMA data, which have been processed as part of the PHANGS--ALMA CO\,(2--1) survey: \\
\noindent ADS/JAO.ALMA\#2012.1.00650.S, \linebreak 
ADS/JAO.ALMA\#2015.1.00925.S, \linebreak 
ADS/JAO.ALMA\#2015.1.00956.S, \linebreak 
ADS/JAO.ALMA\#2017.1.00392.S, \linebreak 
ADS/JAO.ALMA\#2017.1.00886.L, \linebreak 
ADS/JAO.ALMA\#2018.1.01651.S. \linebreak 
ALMA is a partnership of ESO (representing its member states), NSF (USA), and NINS (Japan), together with NRC (Canada), NSC and ASIAA (Taiwan), and KASI (Republic of Korea), in cooperation with the Republic of Chile. The Joint ALMA Observatory is operated by ESO, AUI/NRAO, and NAOJ. The National Radio Astronomy Observatory (NRAO) is a facility of NSF operated under cooperative agreement by Associated Universities, Inc (AUI).

This work is based in part on observations made with NSF's Karl~G.~Jansky Very Large Array 
(VLA; project code:
14A-468, 14B-396, 16A-275, 17A-073, 
18B-184). 
VLA is also operated by NRAO.

This work makes use of data products from the \textit{Wide-field Infrared Survey Explorer (WISE)}, a joint project of the University of California, Los Angeles, and the Jet Propulsion Laboratory/California Institute of Technology, funded by NASA.

This work is based in part on data gathered with the CIS 2.5m Ir\'en\'ee du Pont Telescope and the ESO/MPG 2.2m Telescope at Las Campanas Observatory, Chile.

\vspace{5mm}
\facilities{ALMA, VLA, WISE, Du~Pont, Max~Planck:2.2m,}

\software{\texttt{NumPy} \citep{NumPy_2020},
\texttt{Matplotlib} \citep{Matplotlib_2007},
\texttt{astropy} \citep{Astropy_2013,Astropy_2018,Astropy_2022}
}

}

\bibliography{main}{}

\begin{thebibliography}{}
\expandafter\ifx\csname natexlab\endcsname\relax\def\natexlab#1{#1}\fi
\providecommand{\url}[1]{\href{#1}{#1}}
\providecommand{\dodoi}[1]{doi:~\href{http://doi.org/#1}{\nolinkurl{#1}}}
\providecommand{\doeprint}[1]{\href{http://ascl.net/#1}{\nolinkurl{http://ascl.net/#1}}}
\providecommand{\doarXiv}[1]{\href{https://arxiv.org/abs/#1}{\nolinkurl{https://arxiv.org/abs/#1}}}

\bibitem[{{Astropy Collaboration} {et~al.}(2013){Astropy Collaboration}, {Robitaille}, {Tollerud}, {Greenfield}, {Droettboom}, {Bray}, {Aldcroft}, {Davis}, {Ginsburg}, {Price-Whelan}, {Kerzendorf}, {Conley}, {Crighton}, {Barbary}, {Muna}, {Ferguson}, {Grollier}, {Parikh}, {Nair}, {Unther}, {Deil}, {Woillez}, {Conseil}, {Kramer}, {Turner}, {Singer}, {Fox}, {Weaver}, {Zabalza}, {Edwards}, {Azalee Bostroem}, {Burke}, {Casey}, {Crawford}, {Dencheva}, {Ely}, {Jenness}, {Labrie}, {Lim}, {Pierfederici}, {Pontzen}, {Ptak}, {Refsdal}, {Servillat}, \& {Streicher}}]{Astropy_2013}
{Astropy Collaboration}, {Robitaille}, T.~P., {Tollerud}, E.~J., {et~al.} 2013, \aap, 558, A33, \dodoi{10.1051/0004-6361/201322068}

\bibitem[{{Astropy Collaboration} {et~al.}(2018){Astropy Collaboration}, {Price-Whelan}, {Sip{\H{o}}cz}, {G{\"u}nther}, {Lim}, {Crawford}, {Conseil}, {Shupe}, {Craig}, {Dencheva}, {Ginsburg}, {VanderPlas}, {Bradley}, {P{\'e}rez-Su{\'a}rez}, {de Val-Borro}, {Aldcroft}, {Cruz}, {Robitaille}, {Tollerud}, {Ardelean}, {Babej}, {Bach}, {Bachetti}, {Bakanov}, {Bamford}, {Barentsen}, {Barmby}, {Baumbach}, {Berry}, {Biscani}, {Boquien}, {Bostroem}, {Bouma}, {Brammer}, {Bray}, {Breytenbach}, {Buddelmeijer}, {Burke}, {Calderone}, {Cano Rodr{\'\i}guez}, {Cara}, {Cardoso}, {Cheedella}, {Copin}, {Corrales}, {Crichton}, {D'Avella}, {Deil}, {Depagne}, {Dietrich}, {Donath}, {Droettboom}, {Earl}, {Erben}, {Fabbro}, {Ferreira}, {Finethy}, {Fox}, {Garrison}, {Gibbons}, {Goldstein}, {Gommers}, {Greco}, {Greenfield}, {Groener}, {Grollier}, {Hagen}, {Hirst}, {Homeier}, {Horton}, {Hosseinzadeh}, {Hu}, {Hunkeler}, {Ivezi{\'c}}, {Jain}, {Jenness}, {Kanarek}, {Kendrew}, {Kern}, {Kerzendorf}, {Khvalko}, {King}, {Kirkby}, {Kulkarni},
  {Kumar}, {Lee}, {Lenz}, {Littlefair}, {Ma}, {Macleod}, {Mastropietro}, {McCully}, {Montagnac}, {Morris}, {Mueller}, {Mumford}, {Muna}, {Murphy}, {Nelson}, {Nguyen}, {Ninan}, {N{\"o}the}, {Ogaz}, {Oh}, {Parejko}, {Parley}, {Pascual}, {Patil}, {Patil}, {Plunkett}, {Prochaska}, {Rastogi}, {Reddy Janga}, {Sabater}, {Sakurikar}, {Seifert}, {Sherbert}, {Sherwood-Taylor}, {Shih}, {Sick}, {Silbiger}, {Singanamalla}, {Singer}, {Sladen}, {Sooley}, {Sornarajah}, {Streicher}, {Teuben}, {Thomas}, {Tremblay}, {Turner}, {Terr{\'o}n}, {van Kerkwijk}, {de la Vega}, {Watkins}, {Weaver}, {Whitmore}, {Woillez}, {Zabalza}, \& {Astropy Contributors}}]{Astropy_2018}
{Astropy Collaboration}, {Price-Whelan}, A.~M., {Sip{\H{o}}cz}, B.~M., {et~al.} 2018, \aj, 156, 123, \dodoi{10.3847/1538-3881/aabc4f}

\bibitem[{{Astropy Collaboration} {et~al.}(2022){Astropy Collaboration}, {Price-Whelan}, {Lim}, {Earl}, {Starkman}, {Bradley}, {Shupe}, {Patil}, {Corrales}, {Brasseur}, {N{\"o}the}, {Donath}, {Tollerud}, {Morris}, {Ginsburg}, {Vaher}, {Weaver}, {Tocknell}, {Jamieson}, {van Kerkwijk}, {Robitaille}, {Merry}, {Bachetti}, {G{\"u}nther}, {Aldcroft}, {Alvarado-Montes}, {Archibald}, {B{\'o}di}, {Bapat}, {Barentsen}, {Baz{\'a}n}, {Biswas}, {Boquien}, {Burke}, {Cara}, {Cara}, {Conroy}, {Conseil}, {Craig}, {Cross}, {Cruz}, {D'Eugenio}, {Dencheva}, {Devillepoix}, {Dietrich}, {Eigenbrot}, {Erben}, {Ferreira}, {Foreman-Mackey}, {Fox}, {Freij}, {Garg}, {Geda}, {Glattly}, {Gondhalekar}, {Gordon}, {Grant}, {Greenfield}, {Groener}, {Guest}, {Gurovich}, {Handberg}, {Hart}, {Hatfield-Dodds}, {Homeier}, {Hosseinzadeh}, {Jenness}, {Jones}, {Joseph}, {Kalmbach}, {Karamehmetoglu}, {Ka{\l}uszy{\'n}ski}, {Kelley}, {Kern}, {Kerzendorf}, {Koch}, {Kulumani}, {Lee}, {Ly}, {Ma}, {MacBride}, {Maljaars}, {Muna}, {Murphy}, {Norman},
  {O'Steen}, {Oman}, {Pacifici}, {Pascual}, {Pascual-Granado}, {Patil}, {Perren}, {Pickering}, {Rastogi}, {Roulston}, {Ryan}, {Rykoff}, {Sabater}, {Sakurikar}, {Salgado}, {Sanghi}, {Saunders}, {Savchenko}, {Schwardt}, {Seifert-Eckert}, {Shih}, {Jain}, {Shukla}, {Sick}, {Simpson}, {Singanamalla}, {Singer}, {Singhal}, {Sinha}, {Sip{\H{o}}cz}, {Spitler}, {Stansby}, {Streicher}, {{\v{S}}umak}, {Swinbank}, {Taranu}, {Tewary}, {Tremblay}, {de Val-Borro}, {Van Kooten}, {Vasovi{\'c}}, {Verma}, {de Miranda Cardoso}, {Williams}, {Wilson}, {Winkel}, {Wood-Vasey}, {Xue}, {Yoachim}, {Zhang}, {Zonca}, \& {Astropy Project Contributors}}]{Astropy_2022}
{Astropy Collaboration}, {Price-Whelan}, A.~M., {Lim}, P.~L., {et~al.} 2022, \apj, 935, 167, \dodoi{10.3847/1538-4357/ac7c74}

\bibitem[{{Belfiore} {et~al.}(2023){Belfiore}, {Leroy}, {Sun}, {Barnes}, {Boquien}, {Cao}, {Congiu}, {Dale}, {Egorov}, {Eibensteiner}, {Glover}, {Grasha}, {Groves}, {Klessen}, {Kreckel}, {Neumann}, {Querejeta}, {Sanchez-Blazquez}, {Schinnerer}, \& {Williams}}]{Belfiore_etal_2023}
{Belfiore}, F., {Leroy}, A.~K., {Sun}, J., {et~al.} 2023, \aap, 670, A67, \dodoi{10.1051/0004-6361/202244863}

\bibitem[{{Bigiel} {et~al.}(2008){Bigiel}, {Leroy}, {Walter}, {Brinks}, {de Blok}, {Madore}, \& {Thornley}}]{Bigiel_etal_2008}
{Bigiel}, F., {Leroy}, A., {Walter}, F., {et~al.} 2008, \aj, 136, 2846, \dodoi{10.1088/0004-6256/136/6/2846}

\bibitem[{{Binney} \& {Tremaine}(1987)}]{Binney_Tremaine_1987}
{Binney}, J., \& {Tremaine}, S. 1987, {Galactic dynamics}

\bibitem[{{Brunetti} {et~al.}(2024){Brunetti}, {Wilson}, {He}, {Sun}, {Leroy}, {Rosolowsky}, {Bemis}, {Bigiel}, {Groves}, {Saito}, \& {Schinnerer}}]{2024MNRAS.530..597B}
{Brunetti}, N., {Wilson}, C.~D., {He}, H., {et~al.} 2024, \mnras, 530, 597, \dodoi{10.1093/mnras/stae890}

\bibitem[{{Butler} {et~al.}(2015){Butler}, {Tan}, \& {Van Loo}}]{2015ApJ...805....1B}
{Butler}, M.~J., {Tan}, J.~C., \& {Van Loo}, S. 2015, \apj, 805, 1, \dodoi{10.1088/0004-637X/805/1/1}

\bibitem[{{Chevance} {et~al.}(2020){Chevance}, {Kruijssen}, {Hygate}, {Schruba}, {Longmore}, {Groves}, {Henshaw}, {Herrera}, {Hughes}, {Jeffreson}, {Lang}, {Leroy}, {Meidt}, {Pety}, {Razza}, {Rosolowsky}, {Schinnerer}, {Bigiel}, {Blanc}, {Emsellem}, {Faesi}, {Glover}, {Haydon}, {Ho}, {Kreckel}, {Lee}, {Liu}, {Querejeta}, {Saito}, {Sun}, {Usero}, \& {Utomo}}]{2020MNRAS.493.2872C}
{Chevance}, M., {Kruijssen}, J.~M.~D., {Hygate}, A. P.~S., {et~al.} 2020, \mnras, 493, 2872, \dodoi{10.1093/mnras/stz3525}

\bibitem[{{Chung} {et~al.}(2009){Chung}, {van Gorkom}, {Kenney}, {Crowl}, \& {Vollmer}}]{Chung_etal_2009}
{Chung}, A., {van Gorkom}, J.~H., {Kenney}, J. D.~P., {Crowl}, H., \& {Vollmer}, B. 2009, \aj, 138, 1741, \dodoi{10.1088/0004-6256/138/6/1741}

\bibitem[{{Dowell} {et~al.}(2008){Dowell}, {Buckalew}, \& {Tan}}]{2008AJ....135..823D}
{Dowell}, J.~D., {Buckalew}, B.~A., \& {Tan}, J.~C. 2008, \aj, 135, 823, \dodoi{10.1088/0004-6256/135/3/823}

\bibitem[{{Elmegreen}(1991)}]{1991ApJ...378..139E}
{Elmegreen}, B.~G. 1991, \apj, 378, 139, \dodoi{10.1086/170414}

\bibitem[{{Elmegreen}(1997)}]{1997RMxAC...6..165E}
{Elmegreen}, B.~G. 1997, in Revista Mexicana de Astronomia y Astrofisica Conference Series, Vol.~6, Revista Mexicana de Astronomia y Astrofisica Conference Series, ed. J.~{Franco}, R.~{Terlevich}, \& A.~{Serrano}, 165

\bibitem[{{Gammie} {et~al.}(1991){Gammie}, {Ostriker}, \& {Jog}}]{1991ApJ...378..565G}
{Gammie}, C.~F., {Ostriker}, J.~P., \& {Jog}, C.~J. 1991, \apj, 378, 565, \dodoi{10.1086/170458}

\bibitem[{{Ge} {et~al.}(2023){Ge}, {Wang}, {Duarte-Cabral}, {Pettitt}, {Dobbs}, {S{\'a}nchez-Monge}, {Neralwar}, {Urquhart}, {Colombo}, {Dur{\'a}n-Camacho}, {Beuther}, {Bronfman}, {Rigby}, {Eden}, {Neupane}, {Barnes}, {Henning}, \& {Yang}}]{2023A&A...675A.119G}
{Ge}, Y., {Wang}, K., {Duarte-Cabral}, A., {et~al.} 2023, \aap, 675, A119, \dodoi{10.1051/0004-6361/202245784}

\bibitem[{{Harris} {et~al.}(2020){Harris}, {Millman}, {van der Walt}, {Gommers}, {Virtanen}, {Cournapeau}, {Wieser}, {Taylor}, {Berg}, {Smith}, {Kern}, {Picus}, {Hoyer}, {van Kerkwijk}, {Brett}, {Haldane}, {del R{\'\i}o}, {Wiebe}, {Peterson}, {G{\'e}rard-Marchant}, {Sheppard}, {Reddy}, {Weckesser}, {Abbasi}, {Gohlke}, \& {Oliphant}}]{NumPy_2020}
{Harris}, C.~R., {Millman}, K.~J., {van der Walt}, S.~J., {et~al.} 2020, \nat, 585, 357, \dodoi{10.1038/s41586-020-2649-2}

\bibitem[{{Hunter}(2007)}]{Matplotlib_2007}
{Hunter}, J.~D. 2007, Computing in Science and Engineering, 9, 90, \dodoi{10.1109/MCSE.2007.55}

\bibitem[{{Imara} {et~al.}(2011){Imara}, {Bigiel}, \& {Blitz}}]{2011ApJ...732...79I}
{Imara}, N., {Bigiel}, F., \& {Blitz}, L. 2011, \apj, 732, 79, \dodoi{10.1088/0004-637X/732/2/79}

\bibitem[{{Imara} \& {Blitz}(2011)}]{2011ApJ...732...78I}
{Imara}, N., \& {Blitz}, L. 2011, \apj, 732, 78, \dodoi{10.1088/0004-637X/732/2/78}

\bibitem[{{Kennicutt}(1998)}]{1998ApJ...498..541K}
{Kennicutt}, Robert~C., J. 1998, \apj, 498, 541, \dodoi{10.1086/305588}

\bibitem[{{Kennicutt} \& {Evans}(2012)}]{2012ARA&A..50..531K}
{Kennicutt}, R.~C., \& {Evans}, N.~J. 2012, \araa, 50, 531, \dodoi{10.1146/annurev-astro-081811-125610}

\bibitem[{{Kim} {et~al.}(2022){Kim}, {Chevance}, {Kruijssen}, {Leroy}, {Schruba}, {Barnes}, {Bigiel}, {Blanc}, {Cao}, {Congiu}, {Dale}, {Faesi}, {Glover}, {Grasha}, {Groves}, {Hughes}, {Klessen}, {Kreckel}, {McElroy}, {Pan}, {Pety}, {Querejeta}, {Razza}, {Rosolowsky}, {Saito}, {Schinnerer}, {Sun}, {Tomi{\v{c}}i{\'c}}, {Usero}, \& {Williams}}]{2022MNRAS.516.3006K}
{Kim}, J., {Chevance}, M., {Kruijssen}, J.~M.~D., {et~al.} 2022, \mnras, 516, 3006, \dodoi{10.1093/mnras/stac2339}

\bibitem[{{Koda} \& {Tan}(2023)}]{2023ApJ...959....1K}
{Koda}, J., \& {Tan}, J.~C. 2023, \apj, 959, 1, \dodoi{10.3847/1538-4357/ad05c6}

\bibitem[{{Kruijssen} {et~al.}(2024){Kruijssen}, {Chevance}, {Longmore}, {Ginsburg}, {Ramambason}, \& {Romanelli}}]{2024arXiv240414495K}
{Kruijssen}, J.~M.~D., {Chevance}, M., {Longmore}, S.~N., {et~al.} 2024, arXiv e-prints, arXiv:2404.14495, \dodoi{10.48550/arXiv.2404.14495}

\bibitem[{{Kruijssen} \& {Longmore}(2014)}]{2014MNRAS.439.3239K}
{Kruijssen}, J.~M.~D., \& {Longmore}, S.~N. 2014, \mnras, 439, 3239, \dodoi{10.1093/mnras/stu098}

\bibitem[{{Krumholz} {et~al.}(2009){Krumholz}, {McKee}, \& {Tumlinson}}]{2009ApJ...699..850K}
{Krumholz}, M.~R., {McKee}, C.~F., \& {Tumlinson}, J. 2009, \apj, 699, 850, \dodoi{10.1088/0004-637X/699/1/850}

\bibitem[{{Lang} {et~al.}(2020){Lang}, {Meidt}, {Rosolowsky}, {Nofech}, {Schinnerer}, {Leroy}, {Emsellem}, {Pessa}, {Glover}, {Groves}, {Hughes}, {Kruijssen}, {Querejeta}, {Schruba}, {Bigiel}, {Blanc}, {Chevance}, {Colombo}, {Faesi}, {Henshaw}, {Herrera}, {Liu}, {Pety}, {Puschnig}, {Saito}, {Sun}, \& {Usero}}]{Lang_etal_2020}
{Lang}, P., {Meidt}, S.~E., {Rosolowsky}, E., {et~al.} 2020, \apj, 897, 122, \dodoi{10.3847/1538-4357/ab9953}

\bibitem[{{Lee} {et~al.}(2016){Lee}, {Miville-Desch{\^e}nes}, \& {Murray}}]{2016ApJ...833..229L}
{Lee}, E.~J., {Miville-Desch{\^e}nes}, M.-A., \& {Murray}, N.~W. 2016, \apj, 833, 229, \dodoi{10.3847/1538-4357/833/2/229}

\bibitem[{{Leroy} {et~al.}(2008){Leroy}, {Walter}, {Brinks}, {Bigiel}, {de Blok}, {Madore}, \& {Thornley}}]{Leroy_etal_2008}
{Leroy}, A.~K., {Walter}, F., {Brinks}, E., {et~al.} 2008, \aj, 136, 2782, \dodoi{10.1088/0004-6256/136/6/2782}

\bibitem[{{Leroy} {et~al.}(2012){Leroy}, {Bigiel}, {de Blok}, {Boissier}, {Bolatto}, {Brinks}, {Madore}, {Munoz-Mateos}, {Murphy}, {Sandstrom}, {Schruba}, \& {Walter}}]{leroy_etal_2012}
{Leroy}, A.~K., {Bigiel}, F., {de Blok}, W.~J.~G., {et~al.} 2012, \aj, 144, 3, \dodoi{10.1088/0004-6256/144/1/3}

\bibitem[{{Leroy} {et~al.}(2021){Leroy}, {Schinnerer}, {Hughes}, {Rosolowsky}, {Pety}, {Schruba}, {Usero}, {Blanc}, {Chevance}, {Emsellem}, {Faesi}, {Herrera}, {Liu}, {Meidt}, {Querejeta}, {Saito}, {Sandstrom}, {Sun}, {Williams}, {Anand}, {Barnes}, {Behrens}, {Belfiore}, {Benincasa}, {Be{\v{s}}li{\'c}}, {Bigiel}, {Bolatto}, {den Brok}, {Cao}, {Chandar}, {Chastenet}, {Chiang}, {Congiu}, {Dale}, {Deger}, {Eibensteiner}, {Egorov}, {Garc{\'\i}a-Rodr{\'\i}guez}, {Glover}, {Grasha}, {Henshaw}, {Ho}, {Kepley}, {Kim}, {Klessen}, {Kreckel}, {Koch}, {Kruijssen}, {Larson}, {Lee}, {Lopez}, {Machado}, {Mayker}, {McElroy}, {Murphy}, {Ostriker}, {Pan}, {Pessa}, {Puschnig}, {Razza}, {S{\'a}nchez-Bl{\'a}zquez}, {Santoro}, {Sardone}, {Scheuermann}, {Sliwa}, {Sormani}, {Stuber}, {Thilker}, {Turner}, {Utomo}, {Watkins}, \& {Whitmore}}]{Leroy_etal_2021a}
{Leroy}, A.~K., {Schinnerer}, E., {Hughes}, A., {et~al.} 2021, \apjs, 257, 43, \dodoi{10.3847/1538-4365/ac17f3}

\bibitem[{{Li} {et~al.}(2018){Li}, {Tan}, {Christie}, {Bisbas}, \& {Wu}}]{2018PASJ...70S..56L}
{Li}, Q., {Tan}, J.~C., {Christie}, D., {Bisbas}, T.~G., \& {Wu}, B. 2018, \pasj, 70, S56, \dodoi{10.1093/pasj/psx136}

\bibitem[{{McKee} \& {Ostriker}(2007)}]{2007ARA&A..45..565M}
{McKee}, C.~F., \& {Ostriker}, E.~C. 2007, \araa, 45, 565, \dodoi{10.1146/annurev.astro.45.051806.110602}

\bibitem[{{Poggianti} {et~al.}(2019){Poggianti}, {Gullieuszik}, {Tonnesen}, {Moretti}, {Vulcani}, {Radovich}, {Jaff{\'e}}, {Fritz}, {Bettoni}, {Franchetto}, {Fasano}, {Bellhouse}, \& {Omizzolo}}]{2019MNRAS.482.4466P}
{Poggianti}, B.~M., {Gullieuszik}, M., {Tonnesen}, S., {et~al.} 2019, \mnras, 482, 4466, \dodoi{10.1093/mnras/sty2999}

\bibitem[{{Querejeta} {et~al.}(2021){Querejeta}, {Schinnerer}, {Meidt}, {Sun}, {Leroy}, {Emsellem}, {Klessen}, {Mu{\~n}oz-Mateos}, {Salo}, {Laurikainen}, {Be{\v{s}}li{\'c}}, {Blanc}, {Chevance}, {Dale}, {Eibensteiner}, {Faesi}, {Garc{\'\i}a-Rodr{\'\i}guez}, {Glover}, {Grasha}, {Henshaw}, {Herrera}, {Hughes}, {Kreckel}, {Kruijssen}, {Liu}, {Murphy}, {Pan}, {Pety}, {Razza}, {Rosolowsky}, {Saito}, {Schruba}, {Usero}, {Watkins}, \& {Williams}}]{Querejeta_etal_2021}
{Querejeta}, M., {Schinnerer}, E., {Meidt}, S., {et~al.} 2021, \aap, 656, A133, \dodoi{10.1051/0004-6361/202140695}

\bibitem[{{Rosolowsky} {et~al.}(2003){Rosolowsky}, {Engargiola}, {Plambeck}, \& {Blitz}}]{2003ApJ...599..258R}
{Rosolowsky}, E., {Engargiola}, G., {Plambeck}, R., \& {Blitz}, L. 2003, \apj, 599, 258, \dodoi{10.1086/379166}

\bibitem[{{Schinnerer} \& {Leroy}(2024)}]{2024arXiv240319843S}
{Schinnerer}, E., \& {Leroy}, A.~K. 2024, arXiv e-prints, arXiv:2403.19843, \dodoi{10.48550/arXiv.2403.19843}

\bibitem[{{Schruba} {et~al.}(2010){Schruba}, {Leroy}, {Walter}, {Sandstrom}, \& {Rosolowsky}}]{2010ApJ...722.1699S}
{Schruba}, A., {Leroy}, A.~K., {Walter}, F., {Sandstrom}, K., \& {Rosolowsky}, E. 2010, \apj, 722, 1699, \dodoi{10.1088/0004-637X/722/2/1699}

\bibitem[{{Schruba} {et~al.}(2011){Schruba}, {Leroy}, {Walter}, {Bigiel}, {Brinks}, {de Blok}, {Dumas}, {Kramer}, {Rosolowsky}, {Sandstrom}, {Schuster}, {Usero}, {Weiss}, \& {Wiesemeyer}}]{Schruba_etal_2011}
{Schruba}, A., {Leroy}, A.~K., {Walter}, F., {et~al.} 2011, \aj, 142, 37, \dodoi{10.1088/0004-6256/142/2/37}

\bibitem[{{Silk}(1997)}]{1997ApJ...481..703S}
{Silk}, J. 1997, \apj, 481, 703, \dodoi{10.1086/304073}

\bibitem[{{Sun} {et~al.}(2020{\natexlab{a}}){Sun}, {Leroy}, {Schinnerer}, {Hughes}, {Rosolowsky}, {Querejeta}, {Schruba}, {Liu}, {Saito}, {Herrera}, {Faesi}, {Usero}, {Pety}, {Kruijssen}, {Ostriker}, {Bigiel}, {Blanc}, {Bolatto}, {Boquien}, {Chevance}, {Dale}, {Deger}, {Emsellem}, {Glover}, {Grasha}, {Groves}, {Henshaw}, {Jimenez-Donaire}, {Kim}, {Klessen}, {Kreckel}, {Lee}, {Meidt}, {Sandstrom}, {Sardone}, {Utomo}, \& {Williams}}]{Sun_etal_2020a}
{Sun}, J., {Leroy}, A.~K., {Schinnerer}, E., {et~al.} 2020{\natexlab{a}}, \apjl, 901, L8, \dodoi{10.3847/2041-8213/abb3be}

\bibitem[{{Sun} {et~al.}(2020{\natexlab{b}}){Sun}, {Leroy}, {Ostriker}, {Hughes}, {Rosolowsky}, {Schruba}, {Schinnerer}, {Blanc}, {Faesi}, {Kruijssen}, {Meidt}, {Utomo}, {Bigiel}, {Bolatto}, {Chevance}, {Chiang}, {Dale}, {Emsellem}, {Glover}, {Grasha}, {Henshaw}, {Herrera}, {Jimenez-Donaire}, {Lee}, {Pety}, {Querejeta}, {Saito}, {Sandstrom}, \& {Usero}}]{Sun_etal_2020b}
{Sun}, J., {Leroy}, A.~K., {Ostriker}, E.~C., {et~al.} 2020{\natexlab{b}}, \apj, 892, 148, \dodoi{10.3847/1538-4357/ab781c}

\bibitem[{{Sun} {et~al.}(2022){Sun}, {Leroy}, {Rosolowsky}, {Hughes}, {Schinnerer}, {Schruba}, {Koch}, {Blanc}, {Chiang}, {Groves}, {Liu}, {Meidt}, {Pan}, {Pety}, {Querejeta}, {Saito}, {Sandstrom}, {Sardone}, {Usero}, {Utomo}, {Williams}, {Barnes}, {Benincasa}, {Bigiel}, {Bolatto}, {Boquien}, {Chevance}, {Dale}, {Deger}, {Emsellem}, {Glover}, {Grasha}, {Henshaw}, {Klessen}, {Kreckel}, {Kruijssen}, {Ostriker}, \& {Thilker}}]{Sun_etal_2022}
{Sun}, J., {Leroy}, A.~K., {Rosolowsky}, E., {et~al.} 2022, \aj, 164, 43, \dodoi{10.3847/1538-3881/ac74bd}

\bibitem[{{Sun} {et~al.}(2023){Sun}, {Leroy}, {Ostriker}, {Meidt}, {Rosolowsky}, {Schinnerer}, {Wilson}, {Utomo}, {Belfiore}, {Blanc}, {Emsellem}, {Faesi}, {Groves}, {Hughes}, {Koch}, {Kreckel}, {Liu}, {Pan}, {Pety}, {Querejeta}, {Razza}, {Saito}, {Sardone}, {Usero}, {Williams}, {Bigiel}, {Bolatto}, {Chevance}, {Dale}, {Gensior}, {Glover}, {Grasha}, {Henshaw}, {Jim{\'e}nez-Donaire}, {Klessen}, {Kruijssen}, {Murphy}, {Neumann}, {Teng}, \& {Thilker}}]{Sun_etal_2023}
{Sun}, J., {Leroy}, A.~K., {Ostriker}, E.~C., {et~al.} 2023, \apjl, 945, L19, \dodoi{10.3847/2041-8213/acbd9c}

\bibitem[{{Suwannajak} {et~al.}(2014){Suwannajak}, {Tan}, \& {Leroy}}]{Suwannajak_etal_2014}
{Suwannajak}, C., {Tan}, J.~C., \& {Leroy}, A.~K. 2014, \apj, 787, 68, \dodoi{10.1088/0004-637X/787/1/68}

\bibitem[{{Tan}(2000)}]{Tan_2000}
{Tan}, J.~C. 2000, \apj, 536, 173, \dodoi{10.1086/308905}

\bibitem[{{Tan}(2010)}]{Tan_2010}
---. 2010, \apjl, 710, L88, \dodoi{10.1088/2041-8205/710/1/L88}

\bibitem[{{Tan} {et~al.}(2013){Tan}, {Shaske}, \& {Van Loo}}]{2013IAUS..292...19T}
{Tan}, J.~C., {Shaske}, S.~N., \& {Van Loo}, S. 2013, in IAU Symposium, Vol. 292, Molecular Gas, Dust, and Star Formation in Galaxies, ed. T.~{Wong} \& J.~{Ott}, 19--28, \dodoi{10.1017/S1743921313000173}

\bibitem[{{Tasker} \& {Tan}(2009)}]{2009ApJ...700..358T}
{Tasker}, E.~J., \& {Tan}, J.~C. 2009, \apj, 700, 358, \dodoi{10.1088/0004-637X/700/1/358}

\bibitem[{{Walter} {et~al.}(2008){Walter}, {Brinks}, {de Blok}, {Bigiel}, {Kennicutt}, {Thornley}, \& {Leroy}}]{Walter_etal_2008}
{Walter}, F., {Brinks}, E., {de Blok}, W.~J.~G., {et~al.} 2008, \aj, 136, 2563, \dodoi{10.1088/0004-6256/136/6/2563}

\bibitem[{{Whitmore} {et~al.}(2014){Whitmore}, {Brogan}, {Chandar}, {Evans}, {Hibbard}, {Johnson}, {Leroy}, {Privon}, {Remijan}, \& {Sheth}}]{2014ApJ...795..156W}
{Whitmore}, B.~C., {Brogan}, C., {Chandar}, R., {et~al.} 2014, \apj, 795, 156, \dodoi{10.1088/0004-637X/795/2/156}

\bibitem[{{Wright} {et~al.}(2023){Wright}, {Kounkel}, {Zari}, {Goodwin}, \& {Jeffries}}]{2023ASPC..534..129W}
{Wright}, N.~J., {Kounkel}, M., {Zari}, E., {Goodwin}, S., \& {Jeffries}, R.~D. 2023, in Astronomical Society of the Pacific Conference Series, Vol. 534, Protostars and Planets VII, ed. S.~{Inutsuka}, Y.~{Aikawa}, T.~{Muto}, K.~{Tomida}, \& M.~{Tamura}, 129

\bibitem[{{Wyse} \& {Silk}(1989)}]{1989ApJ...339..700W}
{Wyse}, R. F.~G., \& {Silk}, J. 1989, \apj, 339, 700, \dodoi{10.1086/167329}

\bibitem[{{Zucker} {et~al.}(2018){Zucker}, {Battersby}, \& {Goodman}}]{2018ApJ...864..153Z}
{Zucker}, C., {Battersby}, C., \& {Goodman}, A. 2018, \apj, 864, 153, \dodoi{10.3847/1538-4357/aacc66}

\end{thebibliography}
\bibliographystyle{aasjournal}


\end{document}